\documentclass[twocolumn]{aastex701}

\newcommand{\coto}{\mbox{$\rm{CO}(2-1)$}}
\newcommand{\cott}{\mbox{$\rm{CO}(3-2)$}}
\newcommand{\tcott}{\mbox{$^{13}\rm{CO}(3-2)$}}
\newcommand{\chott}{\mbox{\rm{C}$^{18}$\rm{O}$(3-2)$}}
\newcommand{\kms}{km\,s$^{-1}$}

\newcommand{\asn}{\mbox{AS 205 N}}
\newcommand{\ass}{\mbox{AS 205 S}}
\usepackage{float}
\usepackage{amsmath}
\usepackage{comment}
\usepackage{ulem}
\usepackage{float}
\usepackage{amsmath}
\usepackage{comment}
\usepackage{ulem}
\usepackage{xfrac}
\usepackage{xcolor}
\usepackage{multirow}
\usepackage{pifont}
\usepackage{hyperref}

\begin{document}

\title{A multiwavelength ALMA view of gas and dust in binary protoplanetary system AS 205:\\
Evidence of dust asymmetric distribution}

\correspondingauthor{Nguyen Thi Phuong; Nguyen Tat Thang}
\email{ntphuong02@vnsc.org.vn; thangnguyentat3@gmail.com}

\author[0000-0002-4372-5509]{Nguyen Thi Phuong}
\affiliation{Vietnam National Space Center, Vietnam Academy of Science and Technology, 18 Hoang Quoc Viet, Nghia Do, Hanoi, Vietnam\\}
\affiliation{SAGI Group, Institute For Interdisciplinary Research in Science and Education, 07
Science Avenue, Quy Nhon Nam, Gia Lai, Vietnam\\}
\email[noshow]{ntphuong02@vnsc.org.vn}

\author[0009-0004-1141-0678]{Nguyen Tat Thang}
\email[noshow]{thangnguyentat3@gmail.com}
\affiliation{SAGI Group, Institute For Interdisciplinary Research in Science and Education, 07
Science Avenue, Quy Nhon Nam, Gia Lai, Vietnam\\}
\affiliation{Vietnam National Space Center, Vietnam Academy of Science and Technology, 18 Hoang Quoc Viet, Nghia Do, Hanoi, Vietnam\\}
\affiliation{Department of Physics, University of Rome ``Tor Vergata", Via della Ricerca Scientifica 1, 00133 Rome, Italy}
\affiliation{Department of Astronomy, Faculty of Mathematics, University of Belgrade, Studentski trg 16, 11000 Belgrade, Serbia}
\affiliation{Laboratoire Lagrange, Université Côte d'Azur, Campus Valrose, 28 Avenue Valrose, 06108 Nice Cedex 2, France}

\received{November 05, 2024}
\revised{January 21, 2026}
\accepted{January 22, 2026}

\begin{abstract}

We present Atacama Large Millimeter/Submillimeter Array observations of multi-wavelength dust emissions at 3.1\,mm and 1.3\,mm; along with molecular line emissions of CO(2--1), CO(3--2), \mbox{$^{13}$CO(3--2)}, and C$^{18}$O(3--2) at spatial resolutions of 7--45 AU towards the protoplanetary system AS 205. The dust emissions exhibit two distinct components of AS 205 N and AS 205 S, separated by 1.3 arcsec. While gas kinematics within the dust disk regions are dominated by Keplerian rotation, the more extended gas emission displays complex morphology and kinematics strongly affected by the binary gravitational interaction in the outer regions. The stellar masses of AS 205 N and AS 205 S are estimated at $0.78\pm0.19$ and $1.93\pm0.86$\,M$_\odot$, respectively. Azimuthal variation is observed in the spectral index distribution of both disks. In AS 205 N, the spectral index minimum in the southwest is coincident with the peaks of CO($2-1$), CO($3-2$), and $^{13}$CO($3-2$) integrated intensity and the relative position of its southern counterpart. On the other hand, the spectral index distribution in \ass~exhibits two prominent maxima, with the one in the northeast aligning with the peak of $^{13}$CO($3-2$), and the peak in the south coinciding with local maxima in CO($2-1$) and CO($3-2$) azimuthal profiles. These results suggest a correlation between dust grain size and/or optical depth with the gas distributions. Dust trapping along the spiral arms possibly contributes to the spectral index minima in AS 205 N; however, the observed asymmetry across both disks suggests the involvement of additional mechanisms.
\end{abstract}


\section{Introduction}
Since the seminal discovery of the first exoplanet \citep{Mayor_1995}, the confirmed count of exoplanets has surpassed 6000\footnote{\url{https://exoplanets.nasa.gov/}}, spanning both solitary and multiple systems. Established theories have elucidated the general mechanisms of their formation, wherein planetesimals coalesce from the materials within the protoplanetary disk surrounding their host young forming stars. The study of these protoplanetary materials has progressed significantly over the last decade in light of the advent of Atacama Large Millimeter/Submillimeter Array (ALMA) both in terms of dust (\citealp{2019ApJ...883...71C, 2020ApJ...891...48H, 2021A&A...648A..33M, 2022A&A...664A.137G, 2023ApJ...953...96Z}) and gas properties \citep[][and references therein]{Oberg+etal+2021ApJS..257....1O, Zhang+etal+2021ApJS..257....5Z, Wolfer+etal_2023A&A...670A.154W}, yet most of these detailed studies only concern single disks. 

Planet formation in binary or multiple systems, unlike around single stars, involves a more intricate dynamic. The presence of an additional stellar companion exerts significant perturbations upon the circumstellar disk of the primary star, as it would induce a spiral wake within the disk structure. This phenomenon initiates a torque interaction between the secondary star and the disk, facilitating the transfer of angular momentum from the disk to the secondary star. Consequently, this torque effect leads to the truncation of the primary disk, resulting in the formation of distinct circumstellar disks for both the primary and secondary stars, each confined within their respective Roche lobes \citep{Papaloizou_1977}. Furthermore, theoretical models of disk evolution within young T-Tauri binary systems have revealed the emergence of an encompassing circumbinary disk extending beyond the L2 and L3 Lagrangian points alongside the formation of two inner circumstellar disks. This circumbinary disk exhibits Keplerian rotation around the binary components \citep{Artymowicz_1991}. To fully understand the complex dynamics within this type of disk, it is essential to investigate the materials that make it up. However, research on this subject remains limited, especially regarding its dust component \citep{2021MNRAS.504.2235Z, 2023EPJP..138...25Z}.

AS 205 (or V866 Sco) is a young binary protoplanetary system (SED class II, age $\sim$0.6 Myr, \citealp{Andrews_Huang_2018}), located between the Upper Sco and \mbox{$\rho$-Ophiuchi} star-forming regions. Its two primary components, \mbox{AS 205 N} in the north east and \mbox{AS 205 S} in the south west positions, exhibit a separation of 1.3\arcsec~ \citep[e.g.,][]{Ghez_1993, Reipurth_Zinnecker_1993, McCabe_2006, Andrews_Williams_2007, Salyk_2014, Kurtovic_2018}. \mbox{AS 205 N} is a K5 pre-main-sequence star with the stellar mass of approximately 0.87\,$M_{\rm{\odot}}$ \citep{Andrews_Huang_2018}. \mbox{AS 205 S}, on the other hand, was determined as a spectroscopic binary of spectral types K7 and M0, and masses of 0.74\, M$_{\rm{\odot}}$ and 0.54\, M$_{\rm{\odot}}$, respectively \citep{Eisner_2005}. The distances of \asn~and \ass~were estimated at $132\pm1$ and $142\pm0.3$\,pc, respectively, from their parallax measurements \citep{Gaia+2021}. However, the gas disks encompassing both \asn~and \ass~suggest the existence of gravitational interaction between the two components of the system, making line-of-sight separation of $10\pm4$\,pc unlikely \citep{Salyk_2014, Kurtovic_2018}. Moreover, AS 205 S being an unresolved spectroscopic binary implies that the \citet{Gaia+2021} could not account for the binary motion in their parallax measurements. Consequently, a distance of $132\pm1$\,pc, identical to its northern counterpart, is typically assumed for this disk component.

The investigations conducted by \citet{Salyk_2014} employed ALMA observations of \coto~ and its isotopologues at an angular resolution of 0.6$''$, unveiled non-Keplerian extended emissions originating from the northern component. They are attributed to the effects of tidal interaction, disk wind, or a combination thereof. Similarly, \citet{Kurtovic_2018} scrutinized the spiral-like features observed in the \coto~line emanating from the northern component. Utilizing a higher angular resolution of 0.1$''$, they proposed that these emissions could potentially arise from a flyby scenario. Furthermore, \citet{Weber+etal_2023} later employed the same \coto~line data to estimate the dynamic masses of AS 205 N and AS 205 S, determining them at 0.58\,M$_\odot$ and 0.42\,M$_\odot$, respectively. Additionally, the authors concluded that the stars within the binary system are not gravitationally bound, thereby suggesting the plausibility of a flyby scenario within the system.

This paper presents a comprehensive analysis of ALMA observations of multi-wavelength dust emission at 3.1$\,$mm, and 1.3$\,$mm at a high angular resolution of about 0.05\arcsec~(or $\sim7$AU); as well as molecular line emission of \coto, \cott, \tcott, and \chott~at angular resolutions of 0.12\arcsec -- 0.35\arcsec~\mbox{(or $\sim15-45$\,AU)} in the AS 205 system. The structure of the paper is organized as follows. In Section \ref{sec:obs}, we present the information regarding the observational data and its reduction procedures. Section \ref{sec:results} details the results, with an in-depth analysis of the dust morphology and its properties in Section \ref{sec:dust}; and on the morphology, kinematics, and properties of the gas in Section \ref{sec:gas}. We discuss the results in Section \ref{sec:dis}, and summarize the main points in Section \ref{sec:sum}. 

\section{Observations and data reduction}

The observations of AS 205 were conducted with ALMA Band 7 (2015.1.00168.S, PI: Blake Geoffrey), Band 6 (2016.1.00484.L, PI: Sean Andrews), and Band 3 (2018.1.01198.S, PI: Laura Perez). For the Band 6 data, we used the fiducial data product from the DSHARP collaboration, which was previously reduced and published in \citet{Andrews_Huang_2018} and \citet{Kurtovic_2018}, and is publicly available on their data release website\footnote{\url{https://almascience.eso.org/almadata/lp/DSHARP/}}. The observations and data reduction for the Band 3 and Band 7 data are detailed in the following section.

The Band 3 observations were conducted in six execution blocks (EBs). The correlator was configured in the Time-Division Multiplexing mode, setting up both spectral windows exclusively to target broadband continuum emission. 
The Band 7 observations were carried out in two EBs. One was set to target the molecular lines of \mbox{HCO$^+$\,(4--3)}, HCN\,(4--3) and \cott; while the other covered \tcott, C$^{18}$O\,(3--2), SO\,(8--7) and \mbox{CN\,(3--2)}. Additionally, Band 7 continuum emission was captured in two spectral windows centered at 342.087 GHz and 343.495 GHz, each with a total bandwidth of 1.875 GHz. Detailed information regarding the observing parameters of these observations (i.e., observing time, antenna configuration, number of effective antennas, baseline, total time on source, and calibrators of each EB) is provided in Table \ref{tab:obs}. For both bands, the calibrated measurement sets were acquired upon request from the East Asian ALMA Regional Center (EA-ARC) and subsequently processed using \textsc{CASA} version 6.6.0.

\begin{table*}[htbp]
    \centering
    \caption{Observing parameters of the ALMA observations at Band 3 and Band 7}\label{tab:obs}
    \begin{tabular}{c c c c c c c c}
    \hline
    \textbf{Obs.} & \textbf{Date} & \textbf{Config.} & \textbf{\#Ant} & \textbf{Baseline} & \textbf{TOS} & \multicolumn{2}{c}{\textbf{Calibrator}} \\
    
     & \textbf{(UTC)} & & & \textbf{(m)} & \textbf{(min)} & \textbf{Flux/Bandpass} & \textbf{Phase}\\
    
    (1) & (2) & (3) & (4) & (5) & (6) & (7) & (8)\\
    
    \hline
     & 2018 Oct 11 16:17 & C43-6$^{\color{blue}\rm I}$ & 20 & 15 -- 2516& 43.35 & J1517-2422 & J1617-1941\\ 
     & 2018 Oct 11 17:12 & C43-6$^{\color{blue}\rm I}$ & 26 & 15 -- 2516& 45.42 & J1517-2422 & J1617-1941\\
    Band 3 & 2019 Sep 01 18:50 & C43-6$^{\color{blue}\rm I}$ & 47 & 38 -- 3638 & 45.35 & J1517-2422 & J1617-1941\\ 
     & 2021 Jul 13 01:21 & C43-7$^{\color{blue}\rm I}$ & 43 & 14 -- 3396& 45.33 & J1517-2422 & J1617-1941\\ 
     & 2019 Jun 11 04:28 & C43-9$^{\color{blue}\rm II}$ & 43 & 83 -- 16196 & 42.32 & J1517-2422 & J1617-1941\\
     & 2019 Jun 11 05:49 & C43-9$^{\color{blue}\rm II}$ & 43 & 83 -- 16196& 42.33& J1517-2422 & J1617-1941\\
    
    \hline
    
    Band 7 & 2016 Jun 28 01:17 & C40-4 & 41 & 15 -- 704 & 32.36 & J1517-2422 & J1626-2951\\
     & 2016 Jun 28 02:42 & C40-4 & 41 & 15 -- 704 & 26.32 & J1517-2422 & J1626-2951\\
    
    \hline
    \end{tabular}
    
    \begin{minipage}{15cm}
        \vspace{0.3cm}
        \small  \textbf{Note} (1) Observation band; (2) Date of the observation; (3) Antenna configuration; (4) Number of effective antenna; (5) Baseline; (6) Total time on source; (7) Quasar used for flux and band-pass calibration; (8) Quasar used for phase calibration. 

        $^{\color{blue}\rm I}$ Short-baseline observation.
        $^{\color{blue}\rm II}$ Long-baseline observation.
        \vspace{0.5cm}
    \end{minipage}
\end{table*}

We begin the data reduction process by imaging the continuum emission from each EB. Next, the task \texttt{imfit} is used to determine the peak emission position, which is subsequently used to align the common phase using the tasks \texttt{phaseshift} and \texttt{fixplanets}. We then determine a scale factor from the azimuthally-averaged visibility amplitudes and apply it to each EB to normalize the flux.

Following these initial calibration steps, self-calibration is performed on both datasets. For the Band 3 observations, the self-calibration process was executed in two stages. Initially, self-calibration is run separately for the short-baseline (SB) and long-baseline (LB) data. These two datasets are then combined and re-normalized before undergoing a second round of self-calibration. For all self-calibration instances, both separate and combined datasets, the first four iterations are applied for phase self-calibration only, while the final iteration includes both phase and amplitude self-calibration. Regarding Band 7 observations, five iterations are performed for phase-only self-calibration, followed by a final iteration of both phase and amplitude self-calibration. The resulting self-calibrated visibilities of the continuum data are subsequently imaged using the task \texttt{tclean} with a robust parameter of 0.0.

For our continuum analysis, we limit the data to the high-angular resolution observations from Band 3 and Band 6. The Band 7 continuum data is excluded because its achieved spatial resolution of \mbox{$0.27\arcsec\times 0.20\arcsec$ (P.A.=$-89^\circ$)} at robust = 0.0 is insufficient for studying the detailed dust structures relevant to this work. However, the self-calibration solutions derived from the Band 7 continuum are applied to all molecular line data in this band. The resulting self-calibrated visibilities of the line emissions are then imaged using the task \texttt{tclean} with a robust parameter of 2.0.

Detailed information about the obtained continuum and molecular line data used in this paper is listed in Table \ref{tab:data_red}. 

\label{sec:obs}
\begin{table*}[!htbp]
    \centering
    \caption{Information of the observational data}
    \label{tab:data_red}        
    \begin{tabular}{c c c c c c c c c}    
        \hline   
        \textbf{Emission} & $\boldsymbol{\nu}$ & \textbf{E}$_{\rm \boldsymbol{u}}$ & $\boldsymbol{\Delta}$$\boldsymbol{v}$ & \textbf{Beam} & \textbf{$\boldsymbol{\sigma}_\textbf{\rm I}$} & \textbf{Peak} $\boldsymbol{I_{\nu}}$& \textbf{Proj.code} \\
             
         & (GHz) & (K) & $\left(\frac{\rm km}{\rm s}\right)$ & ($\rm mas\times mas$, $^{\circ}$) & $\left(\frac{\rm \mu Jy}{\rm beam}\right)$ & $\left(\frac{\rm mJy}{\rm beam}\right)$ & & \\    
             
        (1) & (2) & (3) & (4) & (5) & (6) & (7) &  (8)\\
            
        \hline 
            
        3.1\,mm & 101.500 & - & - & $51\times41$, 92.1 & 7 & 2.5 & 2018.1.01198.S\\
             
        1.3\,mm & 233.500 & - & - & $40\times20$, 95.4 & 16 & 6.2 &  2016.1.00484.L \\
             
        \hline
            
        CO(2--1) & 230.545 & 16.6 & 0.35 & $115\times92$, 93.0 & 1418& 75.9  & 2016.1.00484.L \\
             
        CO(3--2) & 345.796 & 33.2 & 0.21 & $324\times225$, 90.7 & 7857 & 885.0 & 2015.1.00168.S \\
             
        $^{13}$CO(3--2) & 330.588 & 31.7 & 0.22 & $474\times297$, 95.8 & 9340 & 406.4 & 2015.1.00168.S\\

        C$^{18}$O(3--2) & 329.330 & 31.6 & 0.22 & $414\times249$, 93.4 & 13830 & 174.5 & 2015.1.00168.S\\
        
        \hline                                  
    \end{tabular} 
    
    \begin{minipage}{15cm}
        \vspace{0.3cm}
        \small  \textbf{Note} (1) Continuum (or Line) emission; (2) Central frequency; (3) Energy of the upper level in line emission; (4) Velocity resolution; (5) Synthetic beam FWHM and its position angle; (6) RMS noise of the emission intensity; (7) Peak Intensity; (8) Project code. 
        \vspace{0.5cm}
    \end{minipage}
\end{table*}

\begin{figure*}[!ht]
    \centering
    \includegraphics[width=0.8\linewidth]{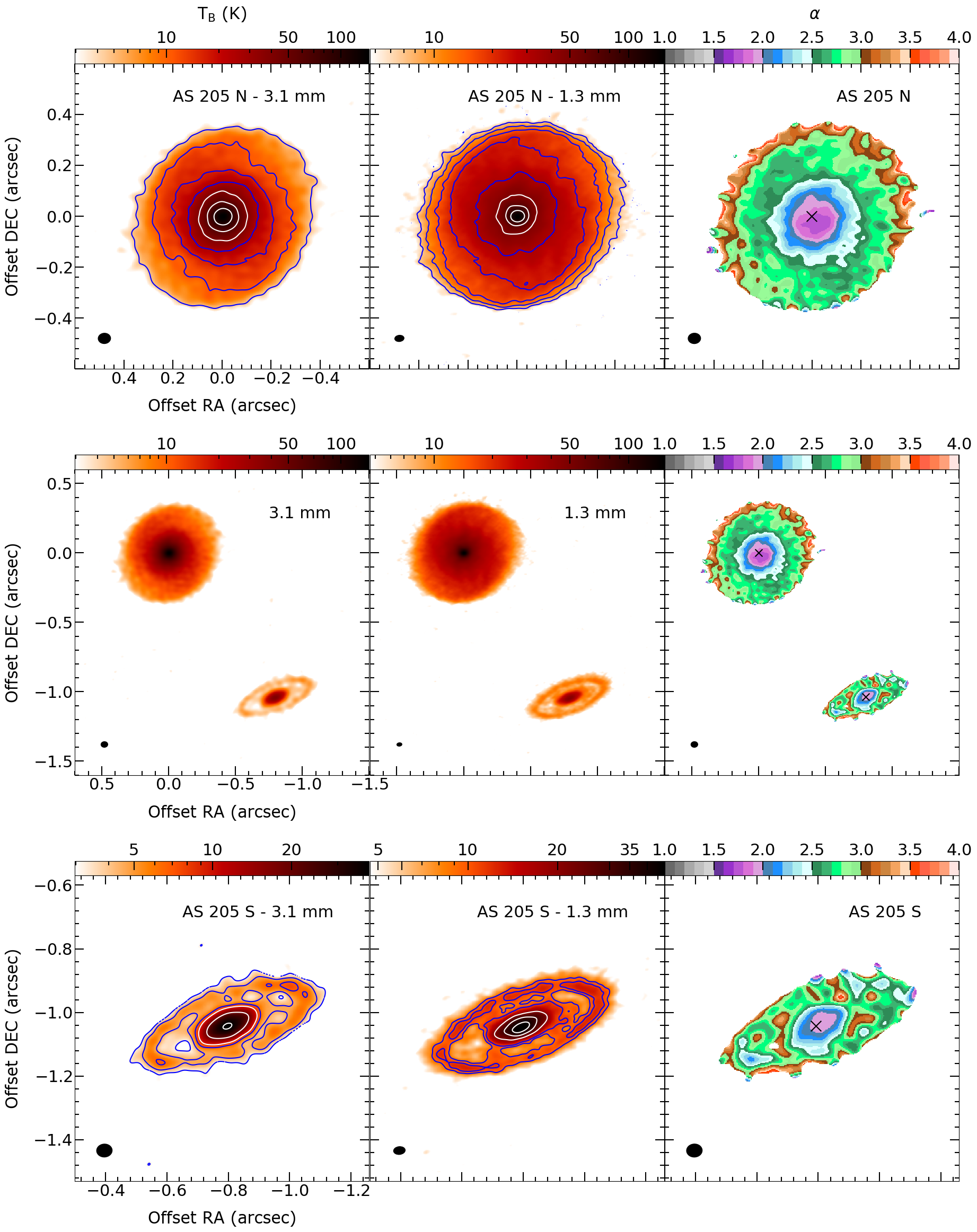}
    \caption{\textit{From left to right}: Brightness temperature distributions of dust continuum emission observed at $\lambda$ = 3.1\,mm and 1.3\,mm; and spectral index derived from the data. The middle panels show the maps of both components, while the top and bottom panels zoom in to \asn~and \ass~components, respectively. The brightness temperature is calculated from the intensity using the Planck function. The beam is shown in the lower left corner with a color bar on the top of each map. The contour levels present in each map are [2.9, 4.6, 10.8, 16.6, 25.5, 39.4, 60.6, 93.3]\,K (\textit{top left}), [4.7, 6.9, 10.2, 15.0, 22.0, 32.4, 47.8, 70.3, 103.6]\,K (\textit{top middle}), [2.5, 4.0, 5.3, 9.0, 18.9, 38.0]\,K (\textit{bottom left}), and [9.0, 11.0, 21.0, 31.0, 41.0]\,K (\textit{bottom middle}). The black crosses in the spectral index maps mark the peak dust emissions of \asn~and \ass.}
    \label{fig:cont-all}
\end{figure*}

%--------------------------------------------------------------------
\section{Results}\label{sec:results}
\subsection{Dust emissions}
\label{sec:dust}
The left and middle panels of Figure \ref{fig:cont-all} present dust continuum images of the protoplanetary system AS 205 observed at $\lambda=3.1$\,mm, and $1.3$\,mm, respectively. The 3.1\,mm observation appears to be largely consistent with the previously observed 1.3\,mm data, with both revealing a pair of disks. The northern disk, AS 205 N, is significantly brighter than its southern counterpart, AS 205 S. While the AS 205 N disk exhibits an asymmetrical distribution in the azimuthal direction, appearing more elongated along the northeast-southwest axis, AS 205 S displays ring and gap structures connected by bridges linking the outer ring to the central region \citep{Kurtovic_2018}.

The two well-separated disks in the 3.1\,mm continuum image are fitted with a 2D Gaussian model using the \textsc{CASA} task \texttt{imfit}. For \mbox{AS 205 N}, we utilize only the SB dataset, which has a beam size of \mbox{$275\times188$\,mas$^2$}, to mitigate any asymmetric features. Additionally, for the outer dust ring of AS 205 S, we fit an ellipse function to the maximum emission points using the routine Markov-chained Monte-Carlo (MCMC) based on the package \textsc{emcee} \citep{2013PASP..125..306F}. Results of the best-fit values are presented in Table \ref{tab:geo}, where they are compared with the values derived from fitting the 1.3\,mm data by \citet{Kurtovic_2018}. 

From Table \ref{tab:geo}, both dust disk components appear to be slightly more compact at 3.1\,mm than previously observed at 1.3\,mm. For AS 205 N, our inclination of $i=17^{\circ}.0\pm2.7$ is in agreement with previous values from the 1.3 mm data, which included a Gaussian fit result of $i=20^{\circ}.1\pm3.3$ and alternative spiral fits yielding $i=15^{\circ}.1_{-3.2}^{+1.9}$ and $i=14^{\circ}.3_{-5.3}^{+1.3}$ \citep{Kurtovic_2018}. Interestingly, our 3.1 mm inclination lies squarely within the range of these previous values. The position angle derived at 3.1\,mm is consistent with the value derived at 1.3 mm by \citet{Kurtovic_2018} within their common uncertainty of $12^\circ$. Moreover, the derived position angle of the disk at both wavelengths has a substantially large uncertainty, likely due to its nearly face-on geometry. For AS 205 S, the inclinations and position angles show negligible differences between the two wavelengths. Furthermore, while the outer ring of AS 205 S exhibits negligible center offset from the whole disk center, it appears slightly more inclined at both wavelengths. For simplicity, we will use the geometrical parameters fitted from 3.1 mm data for all subsequent analyses throughout this paper.

The right panels of Figure \ref{fig:cont-all} show the spectral index map derived from the 3.1\,mm and 1.3\,mm continuum observations. The two continuum images are aligned to match the peak emission of \asn. Following alignment, the peak emission of \ass~in the two images deviates negligibly, by only 4.0\,mas. The 1.3\,mm image is then smoothed out to match the resolution of the 3.1\,mm observation, which is $51\times41$\,mas (PA$\,=\,92.1^\circ$). The pixel-by-pixel spectral index is calculated using the following relation.

\begin{equation}
    \begin{aligned}
        \alpha\,=\,\frac{\mathrm{log}(I_{\nu_{\rm 1.3 mm}}/I_{\nu_{\rm 3.1 mm}})}{\mathrm{log}(\nu_{\rm 1.3 mm}/\nu_{\rm 3.1 mm})},
    \end{aligned}
    \label{equa:spectral_index}
\end{equation}

where $I_{\nu_{\rm 1.3 mm}}$ and $I_{\nu_{\rm 3.1 mm}}$ are the intensity at the observed frequency $\nu_{\rm 1.3 mm}$ and $\nu_{\rm 3.1 mm}$, respectively. The uncertainty of the measured spectral index resulting from the absolute flux calibration uncertainties is 0.17, based on the nominal absolute density flux calibration uncertainty of 5$\%$ for Band 3 and 10$\%$ for Band 6. Note that the flux calibration uncertainty impacts the spectral index uniformly across the image. Thus, as we focus primarily on the spatial variation of the spectral index, the uncertainty mentioned hereafter will consider only the thermal noise uncertainty, unless specifically stated otherwise.

\begin{table*}[!ht]
    \centering
    \caption{Results of the geometrical constraint on the 2 dust disks AS 205 N and AS 205 S}
    \label{tab:geo}
    \begin{tabular}{c c c c c c c}
        \hline
         \textbf{Disk component} & \textbf{R.A.} & \textbf{DEC.} & \textbf{Major axis} & \textbf{Minor axis} & \textbf{Inclination} & \textbf{P.A.} \\
          & (ICRS) & (ICRS) & & & & \\
          & (h:m:s) & (d:m:s) & ( mas ) & ( mas ) & ( $^{\circ}$ ) & ( $^{\circ}$ ) \\
        \hline

        AS 205 N (3.1 mm)$^{\color{blue}\rm I,II}$ & 16:11:31.350 & $-$18:38:26.30 & 315.0$\,\pm\,$3.5 & 301.3$\,\pm\,$2.8 & 17.0$\,\pm\,2.7$ & 99.0$\,\pm\,$12.0\\

        AS 205 S (3.1 mm)$^{\color{blue}\rm II}$ & 16:11:31.295 & $-$18:38:27.34 & 127.6$\,\pm\,$2.9 & 54.3$\,\pm\,$1.4 & 64.8$\,\pm\,$0.9 & 111.5$\,\pm\,$0.9 \\

        AS 205 S - Ring (3.1 mm)$^{\color{blue}\rm II}$ & 16:11:31.295 & $-$18:38:27.34 & 265.5$_{-10.9}^{+12.1}$ & 92.2$_{-4.3}^{+4.4}$ & 69.7$_{-1.3}^{+1.4}$ & 109.6$\,\pm\,$1.5 \\

        \hline
        
        AS 205 N (1.3 mm)$^{\color{blue}\rm I, III}$ & 16:11:31.352 & $-$18:38:26.34 & $414.0\,\pm\,6.0$ & $388.0\,\pm\,6.0$ & $20.1\,\pm\,3.3$ & $114.0$$\,\pm\,$$11.8$ \\

        AS 205 S (1.3 mm)$^{\color{blue}\rm III}$ &  16:11:31.296 & $-$18:38:27.29 & $185.0\,\pm\,6.0$ & 77.0$\,\pm\,$3.0 & $66.3\,\pm\,1.7$ & $109.6$$\,\pm\,$$1.8$ \\

        AS 205 S - Ring (1.3 mm)$^{\color{blue}\rm III}$ & 16:11:31.296 & $-$18:38:27.29 & 266.1$_{-3.1}^{+2.4}$ & 97.6$\,\pm\,$1.6 & 68.4$_{-0.7}^{+0.5}$ & 110.6$_{-0.4}^{+0.5}$ \\
        
        \hline
    \end{tabular}
    
    \begin{minipage}{17cm}
        \vspace{0.3cm}
        \small  \textbf{Note.} 
        
        $^{\color{blue}\rm I}$ Only short-baseline observational data is used for the 2D Gaussian fitting to avoid including asymmetric features.

        $^{\color{blue}\rm II}$ The whole dataset is aligned to the long baseline data taken on June 11$^{\mathrm{th}}$ 2019 of configuration C43-9.

        $^{\color{blue}\rm III}$ \citet{Kurtovic_2018}.
    \end{minipage}
    \vspace{0.5cm}
\end{table*}
        
The derived spectral index generally reaches its minimum value near the center of each disk. For \asn, the minimum spectral index is \mbox{$\alpha\,=\,1.71\,\pm\,0.02$} considering only thermal noise, and \mbox{$\alpha\,=\,1.71\,\pm\,0.17$} including flux calibration error. For AS 205 S, the minimum is \mbox{$\alpha\,=\,1.82\,\pm\,0.06$} with only thermal noise, and \mbox{$\alpha\,=\,1.82\,\pm\,0.18$} including flux calibration uncertainty. The maximum values at $\alpha\sim3.5-4.0$ are typically observed at the outer edge of both disk components. In \asn, the spectral index distribution displays various substructures. Nevertheless, an asymmetry prevails in the northeast-southwest direction, with $\alpha$ typically lower in the southwest. In \ass, a high spectral index is observed in the ``gaps'' between the central region and the outer ring. The outer ring, conversely, exhibits a lower spectral index, whose value reaches $\alpha\,\sim\,2.5$.
% uncertainty 0.05...........0.08

\subsection{Molecular line emissions}
\label{sec:gas}
\subsubsection{Gas morphology}
\begin{figure*}[!ht]
    \centering
    \includegraphics[width=\linewidth]{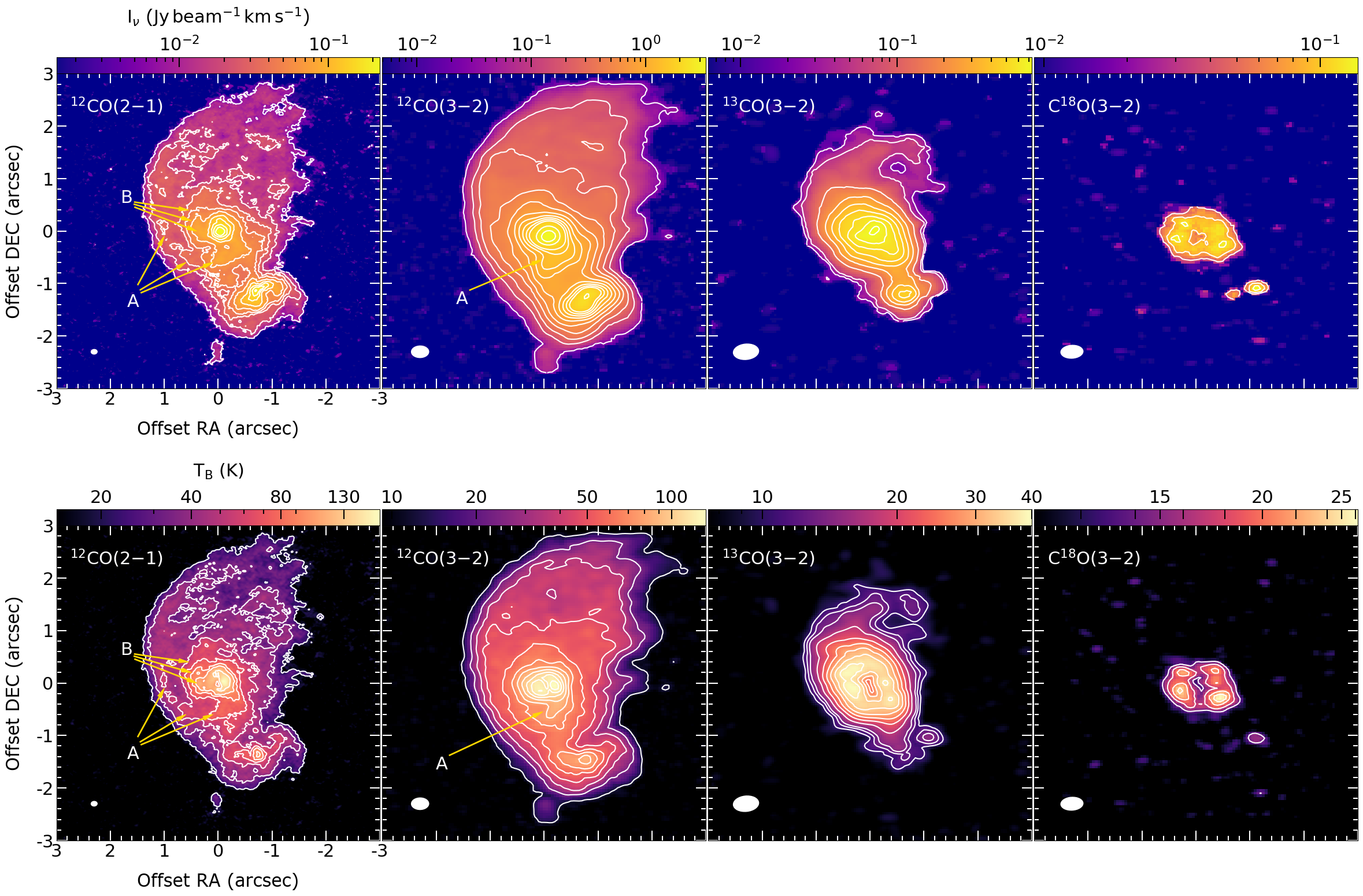}
    \caption{\textit{From left to right:} Integrated intensity maps (\textit{upper panels}) and peak brightness temperature maps (\textit{lower panels}) of \coto, \cott, \tcott, and C$^{18}$O(3--2). The contour levels for the integrated intensity maps are [1, 3, 5, 7, 15, 20, 25, 30]\,$\times \sigma_{^{12}\mathrm{CO}(2-1)}$; [1, 5, 9, 14, 19, 25, 31, 37, 43, 49, 69, 89]\,$\times \sigma_{^{12}\mathrm{CO}(3-2)}$; [0.5, 1, 3, 5, 7, 9, 11, 13, 15]\,$\times \sigma_{^{13}\mathrm{CO}(3-2)}$; and [1, 2, 3]\,$\times \sigma_{\mathrm{C}^{18}\mathrm{O}(3-2)}$, respectively. The contour levels in the peak brightness temperature maps are [5, 10, 15, 20, 25, 30, 40]; [5, 15, 25, 35, 45, 55, 65, 75, 85, 95, 105]; [6, 8, 10, 15, 20, 25, 30, 35, 40]; [5, 7, 9, 11] times of its corresponding RMS noise level, respectively. The beam size and color scale are displayed in the lower left corner and at the top of each panel. In the left panels, A and B denote the spiral-like structures A and B, as mentioned in \citet{Kurtovic_2018}.}
    \label{fig:CO_emissions}
\end{figure*}

Figure \ref{fig:CO_emissions} presents the maps of integrated intensity  (\textit{top}) and peak brightness temperature (\textit{bottom})
for the \coto, \cott, \tcott, and C$^{18}$O(3--2) line emissions. The data for these emission lines are masked below a $3\sigma$ threshold. The brightness temperature is calculated using the Planck function. 

All of the images reveal more extended structures than previously observed in the dust continuum, exhibiting a distinctly elongated morphology in the north-south axis compared to the east-west direction. The line \cott, despite its much lower angular resolution, shares several characteristics with the \coto~emission. These include spiral-like structures emerging from \mbox{AS 205 N} (labeled as A and B in the \coto~and \cott~maps in Figure \ref{fig:CO_emissions}) and a double peak in its peak brightness temperature maps, with the western peak being significantly brighter \citep{Kurtovic_2018}. In contrast, the \tcott~line lacks clear evidence of these spiral-like features, yet displays a unique ``finger-like" structure in the northern hemisphere, which potentially links to the previously noted spiral-like features. Additionally, its peak brightness map distinctly reveals a pronounced intensity dip in the central region around \asn, encircled by a dense and clumpy outer ring. The C$^{18}$O(3--2) maps show much more concentrated emission than the other CO line isotopologues, even comparable to the size of the observed dust disks. Moreover, both its integrated intensity and peak brightness distributions show clumpy ring-like structures in the \asn~disk, while the emission is only weakly detected in the \ass~disk.

This central depression is likely attributed to two major effects, including continuum subtraction when the continuum is optically thick and beam dilution for the substantially large beam size compared to the source \citep{2018ApJ...853..113W, 2025ApJ...994...98F}. The beam dilution effect is expected to have a more pronounced impact on the peak brightness temperature than on the integrated intensity. This explains why the dip is only visible in the peak brightness temperature map of \tcott~but not its integrated map. For much more concentrated \chott, both the continuum subtraction and beam dilution effects are expected to be significantly more severe, leading to the dip being apparent in both its integrated intensity and peak brightness temperature maps.

In addition, the Band 7 observations also include HCO$^+$(4--3) and HCN(4--3) line emissions, which are shown in Appendix \ref{sec:other_lines}.

\subsubsection{Intensity-weighted velocity distributions and Position--Velocity diagrams}
\label{sec:PV}
\begin{figure*}[ht!]
    \centering
    \includegraphics[width=\linewidth]{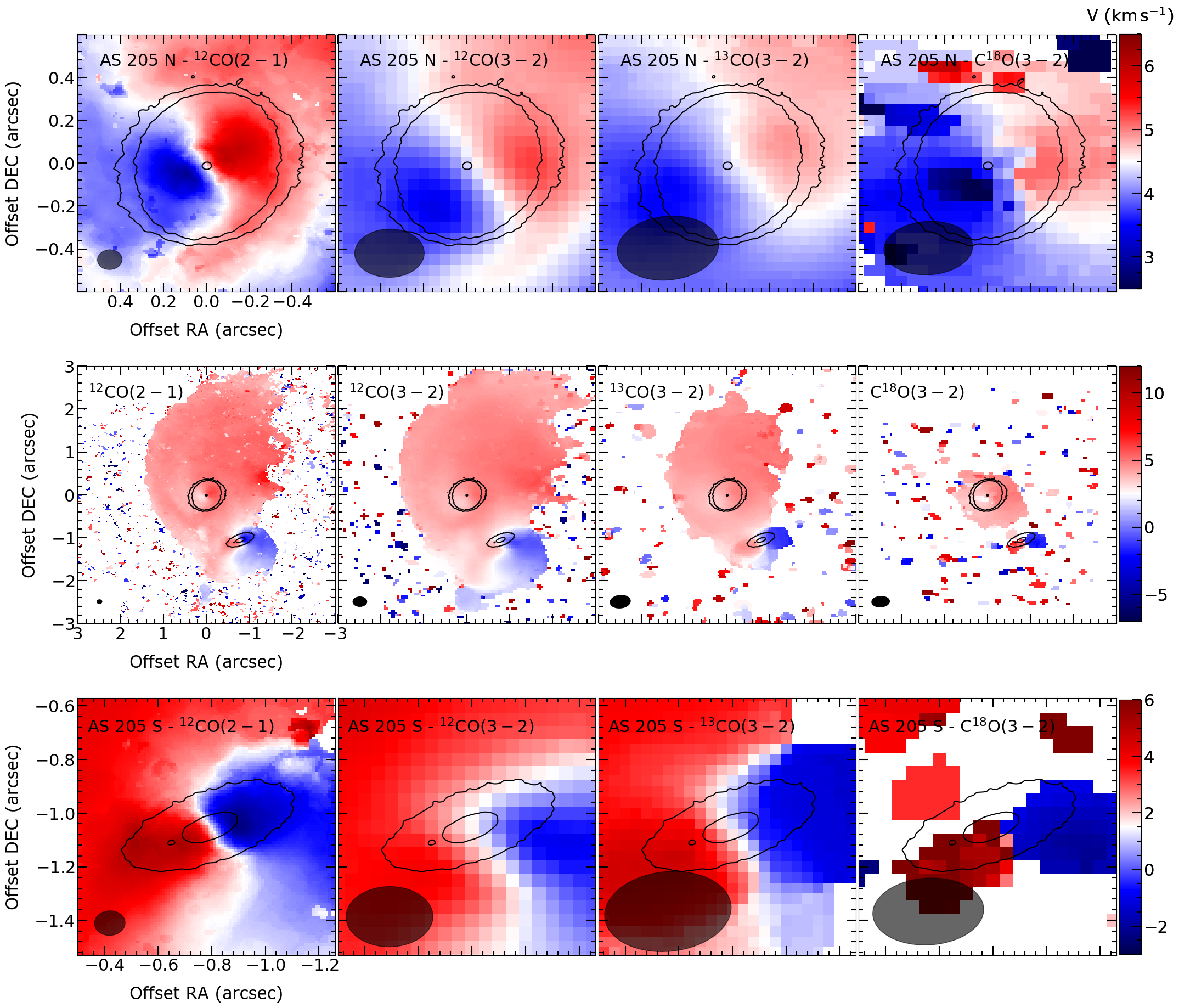}
    \caption{\textit{From left to right}: Mean velocity maps of \coto, \cott, \tcott, and C$^{18}$O(3--2) ~emission of the AS 205 system. The top and bottom panels show the zoom-in images toward the center region of \asn~and \ass, respectively. The black contours show the 1.3\,mm dust emission at the levels of [5, 25, 300]$\sigma$. In each panel, the beam are shown in the lower left corner.}\label{fig:mom1}
\end{figure*}
Figure \ref{fig:mom1} displays maps of the intensity-weighted velocity for the \coto, \cott, \tcott, and C$^{18}$O(3--2) emissions, calculated using a $3\sigma$ threshold. Zooming in on the center of the \asn~component (\textit{top panels}), a clear velocity gradient is visible, showing red-shifted emission in the northwest and blue-shifted emission in the southeast quadrant. Conversely, the \ass~component (\textit{bottom panels}) exhibits the opposite pattern, with red-shifted emission in the southeast and blue-shifted in the northwest quadrant. These features are confined to a small region of the dust disk area. Furthermore, the spiral-like feature A, with the velocity of approximately $4-5$\,\kms, is evident in the \coto~and \cott~maps, but is not visible in the other molecular lines. 

\begin{figure*}[htbp]
    \centering 
    \includegraphics[width=\linewidth]{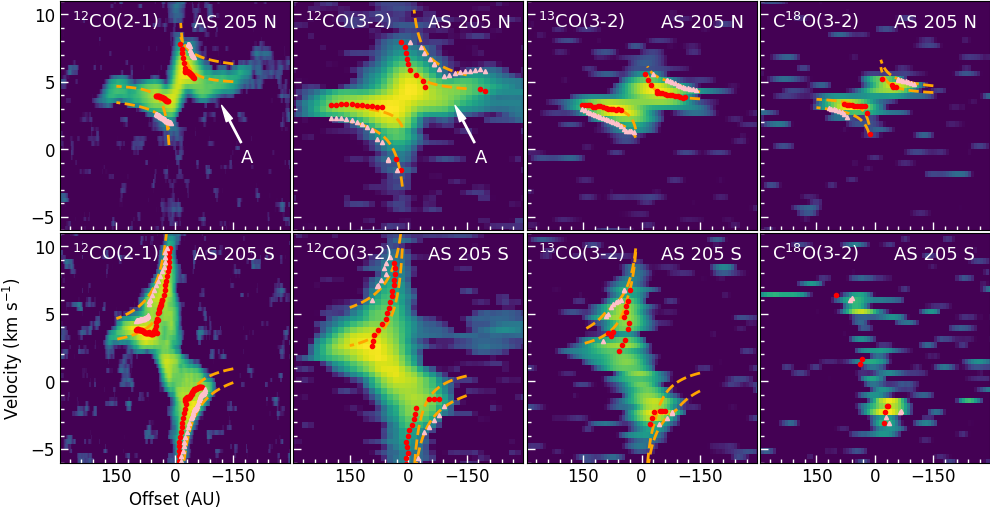} 
    \caption{\textit{From left to right:} PV diagram of four molecular lines \coto, \cott, \tcott, and C$^{18}$O(3--2), cut along the major axis of the observed dust disks \asn~(PA=$99^\circ$, \textit{upper panels}) and \ass~(PA=$110^\circ$, \textit{lower panels}). The extracted points along the edge and ridge are presented in a pink triangle and a red circle, respectively. The orange curves show the fitted results whose parameters are presented in Table \ref{tab:pv_fit_results}. Label A indicates the positions in the diagram of spiral-like structure A, which was shown in Figure \ref{fig:CO_emissions} and originally found in \citet{Kurtovic_2018}.}
    \label{fig:pv-diagrams}
\end{figure*} 

In this section, we examine the Position-Velocity (PV) diagrams extracted from the \coto, \cott, \tcott, and \chott~molecular line emissions. These diagrams were constructed by cutting along the major axes of the observed dust disks in \mbox{AS 205 N} and \mbox{AS 205 S} to estimate the stellar masses of the system.

The PV diagrams exhibit the characteristic butterfly pattern indicative of Keplerian motion within the disks. In the northern component, \asn, the PV diagrams from \coto~and \cott~emissions display spiral-like features in the red-shifted region. In the southern component, \ass, the PV diagrams reveal intensity asymmetry, with the red-shifted region being brighter and more spatially extended than the blue-shifted one. Due to the limited sensitivity of the observations, C$^{18}$O(3--2) emission is only observed around the \asn~disk, while the emission associated with \ass~is only marginally detected.

To determine the stellar mass of \asn~and \ass, we use the public code \textsc{SLAM}\footnote{\url{https://github.com/jinshisai/SLAM}.} \citep{Aso+Sai_2024}, which extracts ridge and edge emission features in each PV diagram for subsequent power-law fitting. The ridge is determined from the peak derived by a 1D Gaussian fit to the emission along the positional axis at each velocity, while the edge is defined by the outermost emission contour above 3$\sigma$ for C$^{18}$O and 4$\sigma$ for the remaining CO isotopologues. The derived positions and velocities $(r,v)$ are then fitted with the following power-law function
\begin{equation}\label{eq:pv}
    \centering 
    V_{rot}=|v-V_{sys}|=v_b \left(\frac{r}{r_b}\right)^{-p}, 
\end{equation}
where $V_{sys}$ is the systemic velocity, $r_b$ is the characteristic radius, and $v_b$ is the velocity measured at $r_b$. In cases where the power-law index $p$ is found to change at the characteristic radius $r_b$, an additional parameter, $dp$, may be introduced, with the corresponding power-law function modified as
\begin{equation}
    \centering
    V_{rot} = v_b \left(\frac{r}{r_b}\right)^{-(p+dp)}
\end{equation} 
Subsequently, under the assumption of Keplerian motion, the stellar mass $M_{\star}$ is derived using the fitted values of $r_b$ and $v_b$, along with the disk inclination $i$ obtained from the dust emission in Section \ref{sec:dust}, with the following relation
\begin{equation}
    \centering
    \frac{v_b}{\sin(i)}=\sqrt{\frac{GM_\star}{r_b}}
\end{equation}
The results of these fits are then presented in Figure \ref{fig:pv-diagrams} and detailed in Table \ref{tab:pv_fit_results}. Note that, for \ass, we do not perform the fit for the PV diagram of C$^{18}$O(3--2) because of its low signal-to-noise ratio (SNR). 

\begin{table*}[!htp]
    \centering 
    \caption{Results of the PV diagram fit using the public code \textsc{SLAM}}
    \label{tab:pv_fit_results}
    \begin{tabular}{|c|l|cccc|ccc|c|}
        \hline
        \multicolumn{2}{|c|}{} & \multicolumn{4}{c|}{AS 205 N} & \multicolumn{3}{c|}{AS 205 S} & Units \\
        \cline{3-9}
        \multicolumn{2}{|c|}{} & $^{12}$CO(2--1) & $^{12}$CO(3--2) & $^{13}$CO(3--2) & C$^{18}$O(3--2) & $^{12}$CO(2--1) & $^{12}$CO(3--2) & $^{13}$CO(3--2) & \\
        \hline
        Edge 
        & $r_b\,^{(1)}$        & $10.59\pm0.04$  & $113.62\pm1.13$  & $82.95\pm4.94$  & $6.50\pm2.47$  &  $57.4\pm22.4$ & $92.16\pm6.86$  &  $90.88\pm1.03$ & AU  \\
        & $v_b\,^{(2)}$        &  $3.91\pm0.04$ &  $1.02\pm0.01$ & $1.45\pm0.05$  & $4.26\pm0.43$  & $4.50\pm1.18$   & $4.11\pm1.17$  &  $3.62\pm0.10$ & $\rm{km\,s}^{-1}$  \\
        & $V_{\rm sys}\,^{(3)}$ & $4.89\pm0.02$  & $3.78\pm0.01$  & $3.53\pm0.01$  & $3.89\pm0.02$   & $2.28\pm0.04$ & $2.21\pm0.79$   & $1.55\pm0.05$  &  $\rm{km\,s}^{-1}$ \\
        & $p\,^{(4)}$ & $0.38\pm0.01$  & $0.58\pm0.02$  &  $0.44\pm0.03$ & $0.53\pm0.08$  & $0.75\pm0.15$  & $0.48\pm0.01$ & $0.46\pm0.07$  & --  \\
        & $dp\,^{(5)}$ & --  & --  &  $0.78\pm0.09$ & --  & --  & -- & $0.49\pm0.20$   & --  \\
        & $M_\star\,^{(6)}$   & $1.56\pm0.03$ &$1.14\pm0.01$ & $1.68 \pm 0.11$  & $1.14\pm0.40$  & $1.57 \pm 0.97$
        & $2.10 \pm 1.10$  &  $1.61 \pm 0.08$ & $\rm{M_\odot}$  \\
        \hline
        Ridge 
        & $r_b\,^{(1)}$        & $13.53\pm0.14$  & $6.03\pm0.41$  & $10.05\pm0.03$   & $13.25\pm0.06$  & $61.8\pm17.9$  & $42.87\pm0.30$  & $35.26\pm3.44$  & AU \\
        & $v_b\,^{(2}$& $5.23\pm0.52$ &  $3.40\pm0.03$ & $2.91\pm0.01$  & $2.39\pm0.24$  & $2.16\pm0.48$  & $3.34\pm0.55$  &  $4.55\pm0.97$ & $\rm{km\,s}^{-1}$  \\
        & $V_{\rm sys}\,^{(3)}$ & $4.83\pm0.01$  & $3.80\pm0.01$   &  $3.49\pm0.01$ & $3.95\pm0.01$  & $2.03\pm0.01$  & $1.53\pm0.18$  & $1.71\pm0.35$  & $\rm{km\,s}^{-1}$  \\
        & $p\,^{(4)}$ & $1.42\pm0.01$  & $0.50\pm0.01$  & $0.46\pm0.13$  & $0.93\pm0.11$  & $0.77\pm0.01$  & $0.87\pm0.13$  & $0.64\pm0.03$  &  -- \\
        & $dp\,^{(5)}$ & --  & --  & $0.46\pm0.31$  &  -- & --   & --   & $0.35\pm0.04$  &  -- \\
        & $M_\star\,^{(6)}$    & $3.56\pm0.7$ &  $0.67 \pm 0.05$  & $0.82 \pm 0.01$  & $0.73\pm0.03$  & $0.39 \pm 0.19$  &  $0.65 \pm 0.19$  &  $0.99 \pm 0.40$ & $\rm{M_\odot}$  \\
        \hline
    \end{tabular}

    \begin{minipage}{18cm}
        \vspace{0.3cm}
        \small  \textbf{Note} (1) Characterized radius; (2) Velocity at characterized radius; (3) Systemic velocity; (4) Power-law index; (5) Power-law index difference between the inner and outer regions of the PV profile; (6) Stellar mass. \\ 
        \vspace{0.5cm}
    \end{minipage}
\end{table*} 

From the stellar mass and systemic velocity values presented in Table \ref{tab:pv_fit_results}, we compute their mean value weighted by the inverse variation from an ideal Keplerian rotating disk, where the power-law index $p=0.5$, as follows
\begin{equation}
    \overline{M_{\star}}\,=\,\frac{\sum_{i}(p_i - 0.5)^{-2}M_{\star,\,i}}{\sum_{i}(p_i - 0.5)^{-2}}
\end{equation}
\begin{equation}
    \overline{V_{los}}\,=\,\frac{\sum_{i}(p_i - 0.5)^{-2}V_{los,\,i}}{\sum_{i}(p_i - 0.5)^{-2}}
\end{equation}

As a result of this procedure, the stellar masses and systemic velocities along the line-of-sight of AS 205 N are derived as $M_N=0.78\,\pm\,0.19\,M_{\odot}$ and $V_{N,\,los}=3.78\pm0.07\,\rm km\,s^{-1}$, and those for AS 205 S are $M_S=1.93\,\pm\,0.86\,M_{\odot}$ and $V_{S,\,los}=2.04\pm0.63\,\rm km\,s^{-1}$. This places the total mass of the system at $M_{N+S}\,=\,2.62\,\pm\,1.05\,M_{\odot}$.

\section{Discussion}\label{sec:dis}
\subsection{Dust disk morphology and dust distribution}
\label{sec:dust_discuss}

As mentioned in Section \ref{sec:results}, the spectral index reaches its minimum near the centers for both AS 205 N and AS 205 S disk components, with values slightly less than 2, indicating possible contribution from free-free emission \citep{2024A&A...684A.134R}. The presence of such emission suggests that accretion onto the central star is still active \citep{2012ApJ...751L..42P}, which is consistent with the relatively young age of approximately $0.6$ Myr \citep{Andrews_Huang_2018} of the system. However, since both of the spectral index values still approach the Rayleigh-Jeans limit ($\alpha\sim2$), the emissions are expected to be dominantly optically thick dust thermal emissions in these central regions of both disk components.

\begin{figure*}
    \centering
    \includegraphics[width=0.88\linewidth]{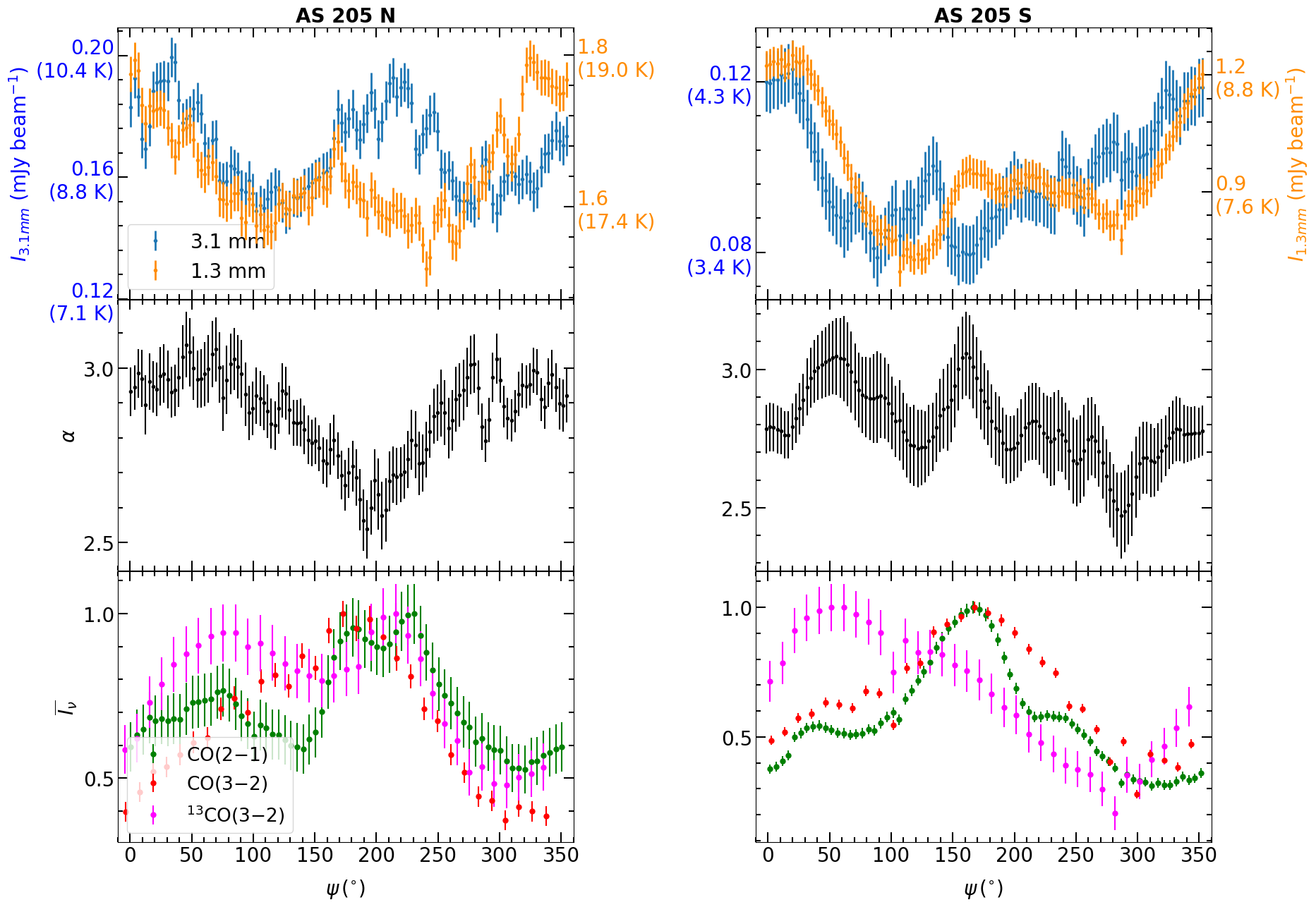}
    \caption{\textit{Top}: azimuthal distribution of dust continuum intensity at 3.1 and 1.3 mm. \textit{Middle}: azimuthal distribution of the spectral index of AS 205 N within the radial distance of $0.2-0.4''$ (\textit{left}) and the outer ring of AS 205 S, which is defined at $0.2-0.4''$ (\textit{right}). \textit{Bottom}: azimuthal distribution of the normalized integrated intensity for CO($2-1$), CO($3-2$), and $^{13}$CO($3-2$), within the radial distance of $0.8''$ in AS 205 N (\textit{left}) and $0.6''$ in AS 205 S (\textit{right}). The azimuthal angle $\psi$ is measured eastward from the north direction.}
    \label{fig:alpha_AS205}  
\end{figure*}

In the outer region of AS 205 N, spanning radii of $0.2 - 0.4''$, and the outer dust ring of AS 205 S, defined over the same radial range, the spectral index largely exceeds 2.5, suggesting that these regions are likely dominantly optically thin, at least marginally. For further investigation, we examine the azimuthal distributions of both the intensity and spectral index along these regions in their respective disk planes. Both of the continuum images at the two wavelengths and the spectral index distribution are smoothed out such that after deprojection, the beam becomes a circular Gaussian with FWHM of 55 and 120 mas in the case of AS 205 N and AS 205 S, respectively. The beam convolutions are performed to avoid the effect of the elliptical beam shape on the azimuthal variations. The profiles are then deprojected using the position angle and inclination derived for each disk as shown in Table \ref{tab:geo}. Subsequently, the azimuthal distributions of intensity and spectral index are presented in the upper and middle panels of Figure \ref{fig:alpha_AS205}, respectively. As a standard convention, the azimuthal angle $\psi$ is defined as the angle measured eastward from the north direction.

Additionally, the lower panels of Figure \ref{fig:alpha_AS205} illustrate the azimuthal profiles of the normalized integrated intensity of CO($2-1$), CO($3-2$), and $^{13}$CO($3-2$) for comparison. These profiles are obtained by first averaging their integrated intensity distribution over the radial ranges of $0.2''$ to $0.8''$ for AS 205 N and $0.6''$ for AS 205 S. This averaging region is substantially larger, which is approximately twice the size of the analyzed dust disk regions, to account for the CO isotopologue data's significantly larger beam size. Moreover, the inner radius of $0.2''$ was excluded to avoid the optically thick emission in the central region with potential contamination from the dust emission. Effectively, these profiles are expected to primarily trace the surrounding gas of the dust disks. Similar to the continuum data, the integrated intensity maps were smoothed prior to generating these azimuthal profiles to ensure circular Gaussian beams are achieved after de-projection for each disk component. %For AS 205 N, the de-projected circular beam FWHM for CO($2-1$), CO($3-2$), and $^{13}$CO($3-2$) integrated intensity distributions are 120, 330, and 500 mas, respectively. In the case of AS 205 S, they are 270, 600, and 1120 mas, respectively. 
The final profiles are then normalized to their respective maximum value for better visualization.

\subsubsection{AS 205 N}
The disk AS 205 N exhibits significant azimuthal asymmetry in its outer region. The intensity profiles at both 3.1 mm and 1.3 mm display a distinct $\mathbf{W}$-shaped feature, characterized by two maxima in the north and southwest directions. While the 3.1 mm profile peaks at $\psi=30^{\circ}\pm10^{\circ}$ and $214^{\circ}\pm12^{\circ}$, the two maxima of the \mbox{1.3\,mm} profile are located at $\psi=339^{\circ}\pm17^{\circ}$ and $167^{\circ}\pm6^{\circ}$.

On the other hand, their corresponding spectral index profile exhibits a distinct dip, dropping by approximately $\Delta\alpha=0.5$ from its surrounding level, centered at $\psi=197^{\circ}\pm10^{\circ}$. Since longer-wavelength observations are more sensitive to larger dust grains, this significant dip in spectral index is consistent with the presence of larger dust grains, which is an indicator of grain growth. However, it can also be attributed to the high optical depth, or a combination of both effects. The position of this spectral index minimum is roughly coincident with the peaks in the southwest directions of both intensity profiles. Furthermore, intriguingly, this direction is also in approximate alignment with the image-plane relative position of its companion disk, AS 205 S, which is in the direction of approximately $217^{\circ}$. This suggests that the asymmetric dust distribution is potentially a result of the dynamical interaction between the two disk components. 
Additionally, a potential local minimum in the spectral index is observed in the azimuthal interval $\psi=[280^{\circ},360^{\circ}]\cup[0^{\circ},70^{\circ}]$ (see Figure \ref{fig:alpha_2} for a better visualization), roughly coinciding with the northern peaks of both 1.3\,mm and 3.0\,mm intensity profiles. However, the decrease in this region is below the noise level, making it difficult to draw a definitive conclusion.

Moreover, in the lower left panel of Figure \ref{fig:alpha_AS205}, the azimuthal profiles of integrated intensity of CO($2-1$), CO($3-2$), and $^{13}$CO($3-2$) around AS 205 N consistently display a twice-the-surrounding-level peak in the south direction, specifically at $\psi=211^{\circ}\pm21^{\circ},\,\psi=180^{\circ}\pm12^{\circ},\,\text{and}\,\psi=211^{\circ}\pm10^{\circ}$, respectively. The azimuthal position of these integrated intensity peaks appears to coincide with the direction of the spectral index minimum in the dust disk, as well as the relative position of the southern counterpart. These correlations imply that the spectral index asymmetric distribution may arise from the gas distribution surrounding the disk, or that both are results of binary interaction. Furthermore, the CO($2-1$) profile is able to resolve two distinct sub-peaks in this narrow range. However, the intensity dip between them is below the noise level, rendering it difficult to conclusively characterize the feature. Additionally, the profiles of CO($2-1$) and $^{13}$CO($3-2$) also exhibit another local maximum in the northeast direction at $\psi=65^{\circ}\pm20^{\circ}\,\text{and}\,\psi=80^{\circ}\pm21^{\circ}$, respectively. 

\subsubsection{Outer ring of AS 205 S}
In the case of the outer ring of AS 205 S, the spectral index distribution reveals two distinct peaks, which are higher than its surroundings by $\Delta\alpha\sim0.4$, centered at $\psi=60^{\circ}\pm15^{\circ}$ and $160^{\circ}\pm7^{\circ}$. These peaks suggest that along these specific directions of the dust ring, either the dust grain size is minimal, the optical thickness is lowest, or a combination of these physical factors is at play. In the lower right panel of Figure \ref{fig:alpha_AS205}, while the integrated intensity profile of $^{13}$CO($3-2$) exhibits a prominent peak in the northeast ($\psi=50^{\circ}\pm19^{\circ}$), those of CO($2-1$) and CO($3-2$) consistently display a peak in the south direction ($\psi=167^{\circ}\pm5^{\circ}$ and $\psi=163^{\circ}\pm7^{\circ}$, respectively). Both of these peaks are approximately twice their surroundings. Similar to its northern companion AS 205 N, this coincidence may indicate that the asymmetric spectral index distribution in this outer ring is a result of the gas distribution surrounding the disk, or they both share the same origin, which is likely the binary interaction. However, in contrast to the spectral index minimum coinciding with the direction of higher gas intensity as in AS 205 N, the disk AS 205 S exhibits spectral index maximum along the direction of higher gas intensity.

\begin{figure}
    \centering
    \includegraphics[width=\linewidth]{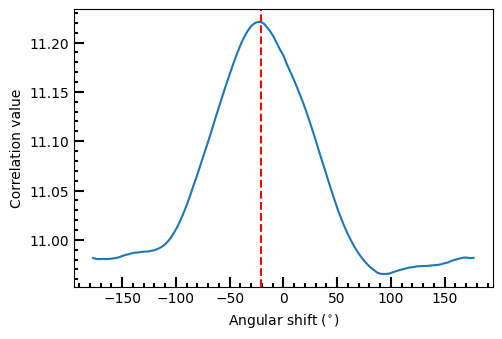}
    \caption{Distribution of circular cross-correlation of the azimuthal intensity profiles of the outer ring of AS 205 S at 3.1 and 1.3 mm, over the range of angular shift [$-180^{\circ}, 180^{\circ}$]. The red dashed line indicates the determined angular shift at $-21^{\circ}$, which corresponds to a maximum cross-correlation value of $11.22$.}
    \label{fig:as205s_corr_shift}
\end{figure}

Moreover, we noted that the intensity profiles at 1.3 mm and 3.1 mm appear to be significantly shifted from each other. To explicitly determine the angular shift between these two azimuthal profiles, we calculate the circular cross-correlation of the two profiles using the Fast Fourier Transform method \citep{Cooley1965} with the function \texttt{fft}\footnote{\url{https://docs.scipy.org/doc/scipy/tutorial/fft.html}} of package \textsc{scipy} in \textsc{Python}. The distribution of the cross-correlation over a range of angular shift is shown in Figure \ref{fig:as205s_corr_shift}. The angular shift is then determined at $-21^{\circ}$, corresponding to a maximum cross-correlation value of $11.22$. The statistical significance of the correlation peak was further assessed using a permutation test with the azimuthal profile at 1.3 mm being randomly shuffled 100,000 times. The permutation test gives a background cross-correlation of 11.08 and a p-value of exactly 0, which indicates that the observed peak correlation is extremely statistically significant. However, assuming an average distance from the center of AS 205 S to its ring is 300 mas (or 0.3$''$), the size of the estimated azimuthal shift projected to the ring is approximately $300 \times \rm \text{sin}(21^{\circ}) \sim 108\, mas$, which is comparable to the projected beam size at 120 mas of this disk. It indicates that this feature is only tentatively detected and still inconclusive.

The observed azimuthal shift may be attributed to Keplerian rotation over the substantial $\sim3$-years temporal separation between the 3.1 mm and 1.3 mm observations. To confirm this particular scenario, we take the combined stellar mass of AS 205 S as $1.93\pm0.86M_{\odot}$, with a distance of the outer ring to the central stars as 40 AU (or 0.3$''$), and the temporal difference of the observations made for the two observations as 3 years. The expected azimuthal shift due to Keplerian motion is $5.8^{\circ}\pm1.4^\circ$, which is significantly smaller than the observed $21^{\circ}$ shift. Namely, the observed azimuthal shift cannot be solely attributed to Keplerian rotation.

The azimuthal shift can also be a result of a possible vertical structure offset of the dust ring. If the ring is optically thick at 1.3 mm, its emission is preferentially sourced from the upper surface layers of the disk. Conversely, the optically thinner 3.1 mm emission is able to penetrate deeper into the disk, originating from regions closer to the midplane. This differential emission origin (disk surface versus deeper layers) effectively translates any vertical offset in the ring's structure into the observed azimuthal offset.

Similar azimuthal shifts have been reported in some crescent-shaped transition disks, such as HD 135344B \citep{2018A&A...619A.161C} and HD 142527 \citep{2019PASJ...71..124S}. In the case of HD 135344B particularly, such a shift has been attributed to the size segregation of dust grains trapped by vortices, as proposed by the same authors. Such a scenario requires that dust grains accumulate and grow around vortices induced by the Rossby Wave Instability (RWI) \citep{1999ApJ...513..805L, 2016ApJ...823...84O}. However, this vortex-trapping scenario is likely not applicable in the case of \ass, where the brightest intensity peak along the observed ring is only about 1.5 times the surrounding emission, which is insufficiently strong to indicate a significant dust trap formed by RWI-induced vortices. Instead, the clumpiness observed in the ring is most likely an artifact resulting from the limited SNR of the data, rather than an intrinsic physical feature of the disk.
 
In summary, the azimuthal intensity shift observed between the two wavelengths is most naturally explained as an intrinsic consequence of a vertical structural offset in the disk. A minor contribution from the Keplerian rotation and temporal delay between the observations, however, cannot be ruled out.

\subsection{Possible origins of the azimuthal asymmetry in the spectral index distribution}
\label{sec:discuss:ori_asym}
In Section \ref{sec:dust_discuss}, we showed that both dust disk components in AS 205 exhibit significant azimuthal asymmetry in their spectral index distributions, wherein the position of the spectral index extrema (minimum for AS 205 N and maximum for AS 205 S) appears to coincide with the direction of the peaks in the CO isotopologues integrated intensity profiles for their respective disks. Moreover, in AS 205 N, particularly, the azimuthal direction of the spectral index minimum is also aligned with the relative position of its southern companion. These features collectively suggest that the spectral index asymmetric distribution is likely a result of the binary interaction between the two components. Here, we further explore in detail the potential factors and mechanisms that can produce this observed asymmetry.

\begin{figure}
    \centering
    \includegraphics[width=0.85\linewidth]{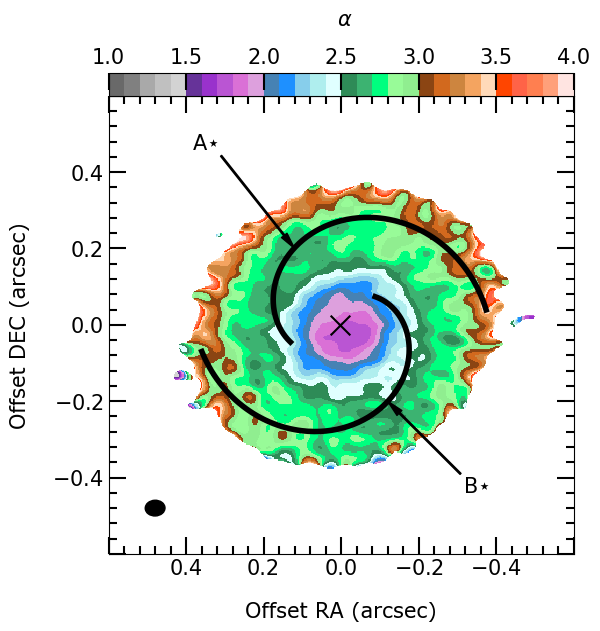}
    \caption{Map of the spectral index in AS 205 N, overlay with black curves representing the spiral arms fitted with Archimedean spiral model derived in \citet{Kurtovic_2018}. A$\star$ and B$\star$ are the denotations of the spiral arms.}
    \label{fig:alpha_spiral}
\end{figure}

\textit{Dust trapping along spiral arms?} AS 205 N was revealed to host a prominent two-armed ($m=2$) spiral structure \citep{Kurtovic_2018}. These spirals likely originate from gas density waves, which are potentially generated by the tidal interaction of AS 205 N with its stellar companion in AS 205 S \citep{1979ApJ...233..857G, 2002ApJ...565.1257T}. Such density waves are expected to establish local pressure maxima that can effectively trap large dust grains as aerodynamic drag causes them to lose angular momentum and drift toward the arms \citep{{Birnstiel+etal_2016SSRv..205...41B}}.
The observed spiral morphology is likely the cause of the characteristic $\mathbf{W}$-shaped azimuthal intensity profiles seen in the top left panel of Figure \ref{fig:alpha_AS205}. Notably, the intensity peak at $\psi=215^{\circ}\pm11^{\circ}$ in the 3.1 mm profile, which likely corresponds to the over-density feature in spiral arm B$^\star$ as denoted in Figure \ref{fig:alpha_spiral}, coincides with the significant minimum in the spectral index distribution. Similarly, the intensity peaks in the north direction of both intensity profiles, which are potentially a manifestation of spiral arm A$^\star$ as shown in Figure \ref{fig:alpha_spiral}, are spatially coincident with the tentative decrease below the noise level found within the azimuthal interval $\psi=[280^{\circ},360^{\circ}]\cup[0^{\circ},70^{\circ}]$. These correlations suggest that the asymmetrical distribution of the spectral index in AS 205 N may be a direct indication of dust over-density and/or dust size-dependent migration towards the spiral arms. However, while the amplitude of the spectral index minimum in the southwest is significantly larger than its tentative counterpart in the north direction, their corresponding two peaks in the intensity profile show a non-significant difference at both observed wavelengths. This contrast implies that, besides dust trapping along the spiral arms, there must be at least one additional factor at play that causes the spectral index decrease in the southwest direction. This mechanism may have also created a north-southwest spectral index asymmetry, leading to an increase in the spectral index toward the north and a decrease toward the southwest. The increase balances the spiral-induced spectral index dip when averaged out in the north, while the decrease amplifies the spectral index decrease in the southwest.

\textit{Connecting gas flow?}  
The CO($2-1$), CO($3-2$), and $^{13}$CO($3-2$) integrated intensity peaks corresponding to the spectral index minimum in AS 205 N and spectral index maximum in AS 205 S are likely part of a larger structure, possibly a connecting gas flow between the two binary components. To investigate whether this presumed connecting gas flow genuinely affects the dust dynamics and subsequently, dust distribution within the dust disk regions, we calculate the Roche lobe radii of the two binary components as follows

\begin{equation}
    \begin{aligned}
        \frac{r_{1}}{A}\,=\,\frac{0.49\,q^{2/3}}{0.6\,q^{2/3}\,+\,\mathrm{ln}(1+q^{1/3})},
    \end{aligned}
\end{equation}
where $r_{1}$ is the radius of the sphere whose volume approximates the Roche lobe of mass $M_{1}$, A is the orbital separation of the system, and $q\,=\,M_{1}/M_{2}$ is the mass ratio of the object of interest $M_{1}$ and its companion $M_{2}$. A$\,=\,1.3''$ as the orbital separation in the image plane is taken as the lower limit of the intrinsic one. From the PV diagram fitting analyses in Sections \ref{sec:PV}, the stellar masses of AS 205 N and AS 205 S were derived at $0.78\,\pm\,0.19\,\mathrm{M}_\odot$ and $1.93\,\pm\,0.86\,\mathrm{M}_\odot$, respectively. Consequently, the approximate lower limits of the Roche lobe radii for AS 205 N and AS 205 S are $0.41''\,\pm\,0.07''$ and $0.59''\,\pm\,0.07''$, respectively. While it is still ambiguous with this result to determine the connection between the spectral index distribution in AS 205 N and its surrounding connecting gas flow, the lower limit of Roche lobe radius in AS 205 S is well-larger than its observed dust-disk radius, implying that dust grains within this disk are unlikely to be affected by the gas flow connecting the two disk components. 

\textit{Differential dust radial drift?} Even within the Roche lobe, the gravitational influence exerted on each disk by the central star of its companion is still present. This gravitational effect is expected to play a crucial role in dust radial drift, which in turn, dominates over dust fragmentation in limiting grain size in the disk outer regions \citep{2012A&A...539A.148B}. As a result, the dust distribution may vary azimuthally due to the uneven gravitational pulls from the companion's host star, leading to differences in the radial drift of dust. 

In the Epstein regime, assuming the Stokes number $St<1$, the grain size limit set by the inward dust radial drift following \citet{2012A&A...539A.148B} can be expressed as
\begin{equation}
    \begin{aligned}
        a_{\rm drift}\,=\,f_{d}\frac{2}{\pi}\frac{\Sigma_{d}}{\rho_s}\frac{v_{K}^{2}}{c_{s}^{2}}\gamma^{-1}
    \end{aligned}
    \label{equa:a_drift}
\end{equation}
where $f_{d}$ is a calibration factor, $\Sigma_{d}$ is the dust surface density, $\rho_s$ is the dust particle bulk density, $v_{K}$ is the Keplerian velocity, $c_s$ is the sound speed, and $\gamma=|d\,\mathrm{log}\,P/d\,\mathrm{log}\,R|$ is the absolute value of the power-law index of the gas pressure profile.

The companion's gravity most notably affects the Keplerian velocity $v_K$ within the host disk. On the disk's far side relative to the companion, $v_K$ increases due to the net gravitational acceleration from both bodies. Conversely, on the closer side, $v_K$ decreases as the companion's gravity partially counteracts the host star's dominant gravitational pull\footnote{This may seem like an abusive way of using the term Keplerian velocity $v_{\rm K}$ when referring to different azimuthal position, yet at the same radial distance from the central star. However, since both the gas and dust experience the gravitational forces identically, all factors related to $v_{\rm K}$ in the equations of motion simply scale with the gravitational forces. Thus, to simplify, we only discuss $v_{\rm K}$.}. For this discussion, we neglect the companion's gravitational effect on the gas pressure gradient and dust density on the disk for simplicity. This allows us to focus on how the maximum grain size, determined by the radial drift limit in the outer regions, varies azimuthally. Specifically, the grain size on the near side of the companion disk is expected to be smaller than that on the far side. For this scenario to be validated, the relative 3D positioning of the two dust disks is needed. Interestingly, the position angles of both disk components are well-aligned with the axis connecting the centers of these disks. This implies that even if the line-of-sight separation of the two dust disks exceeds their separation in the plane of sky, the northeast and southwest directions of each disk can still correspond to either the nearest or farthest side on the disk relative to the companion. Thus, to attribute the observed northeast-southwest dust asymmetry in the disk AS 205 N to azimuthal variations in dust radial drift, the northeast side of this disk must be closer to its companion. Since the northeast side of the AS 205 N disk is known to be closer to the observer \citep{Weber+etal_2023}, this side could only be nearer to AS 205 S if the AS 205 N disk is located farther along the line of sight, with a separation from AS 205 S that is substantially larger than the plane-of-sky one, by a factor of at least $\text{tan}^{-1}(i_{AS\,205\,N})\,\sim\,3$.

However, we emphasize that, in the drift-limited regime, large grains are removed when their drift timescale is shorter than the time required for their formation. Consequently, for a specific grain size to remain in the outer regions of the disk, its radial drift timescale must exceed the grain growth timescale, which can be approximated as 
\begin{equation}
    \begin{aligned}
        t_{\rm grow}\,=\,\frac{1}{\epsilon\,\Omega}
    \end{aligned}
\end{equation}
where $\epsilon$ is the dust-to-gas ratio, and $\Omega$ is the Keplerian frequency (e.g., \citealp{2008A&A...480..859B, 2012A&A...539A.148B}). Therefore, in the typical condition of dust-to-gas ratio $\epsilon=0.01$, the grain growth timescale is 100 times the dynamical timescale. This significantly protracted timescale suggests it is unlikely that the grains can grow rapidly enough to establish the observed dust size asymmetric distribution. Nevertheless, we stress that the possibility of a high dust-to-gas ratio in the disk cannot be ruled out, which may indicate a faster grain growth process and thus permit the azimuthal grain size asymmetry.

In conclusion, while dust trapping along the spiral arms is expected to contribute to the spectral index asymmetry observed in AS 205 N, there must be at least one additional factor responsible for the strong spectral index minimum in its southwest region. The similar spectral index asymmetry observed in AS 205 S, which currently lacks a detailed explanation, further suggests that additional mechanism(s) are at play to drive the observed spectral index asymmetry in both dust disks. However, we have yet to identify such mechanism(s).

\subsection{Gravitationally bound binary or Hyperbolic flyby?} 
Gas emission in the AS 205 binary system shows strong connections and extended structures between the two components. While the observations do not reveal a circumbinary disk or ring, the gas emission exhibits a significant asymmetry, specifically with extended non-Keplerian gas at the velocity of approximately $5-7\,\rm km\,s^{-1}$ concentrated in the northern hemisphere. This non-Keplerian gas was first reported by \citet{Pontoppidan+etal_2011Msngr.143...32P} and \citet{Brown+etal_2013ApJ...770...94B} using spectrographic observations. Subsequent high-resolution mapping using ALMA by \citet{Salyk_2014} and \citet{Kurtovic_2018} confirmed this motion. The nature of this motion is currently under debate, with proposed mechanisms including a molecular disk wind or outflow originating from the disk, or a tidal interaction scenario, such as a hyperbolic flyby.

A critical question related to this problem is whether the AS 205 system is gravitationally bound or undergoing a hyperbolic flyby, which has been addressed previously by \citet{Weber+etal_2023}. By comparing the disk components’ relative velocity with the system’s escape velocity, the authors proposed that the AS 205 system is undergoing a hyperbolic flyby, independent of the line-of-sight distance of the system. However, their analysis is also strongly dependent on the constrained total stellar mass and plane-of-sky velocity of the system, which are substantially different from our result as estimated in Section \ref{sec:PV}. Therefore, as an adaptation to their work, we re-perform the analysis as follows.

\begin{figure}
    \centering
    \includegraphics[width=0.96\linewidth]{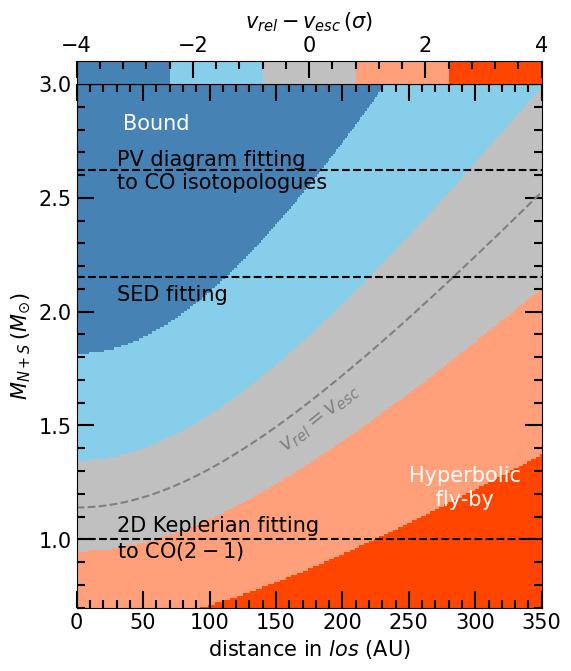}
    \caption{Adapted from Figure 11 in \citet{Weber+etal_2023}, comparison of relative velocity ($v_{rel}$) between AS 205 N and AS 205 S
to the system’s escape velocity $v_{esc}$, depending on its total mass and the components’ separation along the line-of-sight (los). The velocity difference is given as a multiple of the uncertainty of the relative velocity $\sigma=0.37\,\rm km\,s^{-1}$. The grey dashed line shows where the velocities are equal. Parameter space in the top-left is for a bound system, while the bottom-right is unbound, indicating a hyperbolic flyby. The lower dashed line marks the total stellar mass estimate using 2D Keplerian fitting to the velocity map of CO($2-1$) in \citet{Weber+etal_2023}, the middle dashed line marks the result done with spectroscopic energy distribution (SED) fitting from \citet{Eisner_2005} and \citet{Andrews_Huang_2018}, and the upper dashed line marks our result using PV diagram fitting to different CO isotopologues.}
    \label{fig:dynamic_state}
\end{figure}

The absolute value of the disk components' relative proper motion in the image plane was determined at $2.90\pm0.21\,\rm km\,s^{-1}$ \citep{Weber+etal_2023}, based on the values of proper motion of AS 205 N and AS 205 S derived in \citet{Gaia+2021}. On another hand, the velocities along the line-of-sight of AS 205 N and AS 205 S are $V_{N,los}=3.78\pm0.07\rm\, km\,s^{-1}$ and $V_{S,los}=2.04\pm0.63\,\rm km\,s^{-1}$, respectively, which gives their relative velocity along this direction as \mbox{$1.74\pm0.63\rm \,km\,s^{-1}$}. The absolute relative velocity is then computed at $3.38\pm0.37\rm \,km\,s^{-1}$. Subsequently, we show the dependence of the dynamical state of the system on the stellar mass and line-of-sight distance between the binary components in Figure \ref{fig:dynamic_state}. 

Our estimated total mass $M_{N+S}=2.62\pm1.05\,M_{\odot}$ is more than twice the value $1.0\pm0.11\,M_{\odot}$ found in \citet{Weber+etal_2023}, yet consistent with the value $2.15\,M_{\odot}$ derived by \citet{Eisner_2005} and \citet{Andrews_Huang_2018}. The stellar mass in \citet{Eisner_2005} and \citet{Andrews_Huang_2018} was derived using spectroscopic data, which is entirely independent of this work. Whereas \citet{Weber+etal_2023} constrained the stellar mass by fitting a 2D Keplerian model to the CO($2-1$) velocity map of the same data used in this paper. Comparing with this result specifically, our derived stellar mass of AS 205 N at $0.78\pm0.19\,M_{\odot}$, with PV analysis in Section \ref{sec:PV}, is in good agreement with theirs at $0.58\pm0.05\,M_{\odot}$. On the other hand, our constrained stellar mass of AS 205 S, $1.93\pm0.86\,M_{\odot}$, is significantly larger than the value $0.42\pm0.06\,M_{\odot}$ found in \citet{Weber+etal_2023}. Their result, however, is similar to the value we obtained from fitting the ridge of our PV diagrams (Table \ref{tab:pv_fit_results}), where substantial deviation from Keplerian rotation is expected, as indicated by the power-law index of $p\gg0.5$. This suggests that their 2D Keplerian fit is likely biased towards the velocity of the brighter, more strongly-emitting gas in the central velocity regions, which is also strongly non-Keplerian. Therefore, their derived stellar mass may be prone to errors caused by these strong non-Keplerian components in the disk. 
We further note that, although our estimated stellar mass of AS 205 S is consistent with the result of $1.28\,M_{\odot}$ in \citet{Eisner_2005}, it is associated with a large uncertainty. Moreover, the spectroscopic modeling approach, as in \citet{Eisner_2005}, is highly dependent on the assumed stellar model. On the other hand, kinematic analyses like that of \citet{Weber+etal_2023} and our own work are complicated by the strongly perturbed and non-Keplerian nature of the surrounding gas. Therefore, future work is still needed to better constrain the stellar mass of this complex and strongly perturbed system. 

As a result, from Figure \ref{fig:dynamic_state}, the dynamical state of the system AS 205 is again set back to uncertainty following the unknown distance along the line-of-sight of the binary components and a large value of the derived total stellar mass.

\section{Summary}
\label{sec:sum} 
This paper presents the results of high angular resolution ALMA observations of dust emissions at 3.1\,mm and 1.3\,mm (at the resolution of 7 AU) and molecular line emission of CO(2--1), CO(3--2), $^{13}$CO(3--2), and C$^{18}$O(3--2) (at the resolution of 15--45 AU) in a multiple stellar system, AS 205. Our main results are summarized as follows.

\begin{itemize}

\item The dust emission resolves two $\sim 0.4\arcsec$ disks of \asn~and \ass~separated by $1.3\arcsec$, while the more extended gas emission displays complex morphology and kinematics, indicating strong interaction. This gas includes non-Keplerian motion extending to $\sim 2\arcsec$ in the northern outskirt, as revealed by \coto, \cott, and \tcott~ lines, linked by spiral features connecting \asn~to the northern gas.

\item The \asn~dust disk displays an azimuthal asymmetry in its spectral index distribution, with the minimum value centered in the southwest quadrant. Furthermore, the dust emission from the outer ring of \ass~is tentatively found to be azimuthally offset by $21^{\circ}$ between the 3.1 mm and 1.3 mm observations.

\item PV analysis of CO isotopologues determines the stellar masses of the system to be $0.78 \pm 0.19\,\text{M}_\odot$ for \asn~and $1.93 \pm 0.86\,\text{M}_\odot$ for the \ass.

\item In \asn, the minimum spectral index position is coincident with the relative position of its southern counterpart and the peak integrated intensity of \coto, \cott, and \tcott~profiles. In \ass, the spectral index distribution shows two pronounced peaks, with the one in the northeast aligns with that of \tcott~profile, and the southern one consistently coincides with both peaks in \coto~and \cott~profiles.

\item Although the dust trapped in the spiral arms likely contributes to the spectral index minima in \asn, the observed asymmetry across both disks suggests the involvement of additional mechanisms.

\item Our comparison analysis of the disk components’ relative velocity with the system’s escape velocity suggests that the mystery regarding the dynamical state of AS 205 remains an open question, unlike the concluded flyby scenario as in \citet{Weber+etal_2023}.

\end{itemize}

\section*{Acknowledgment} 
We thank the anonymous reviewer for very constructive comments that helped us improve the quality of our manuscript; ALMA East Asian ARC, especially Dr. Atsushi Miyazaki, for their help in calibrating the data from the observations of projects 2015.1.00168.S and 2018.1.01198.S; and Dr. Akimasa Kataoka (NAOJ) and Dr. Kiyoaki Doi (MPIA) for their helpful discussions regarding the dust asymmetric distribution.  This research was funded by Vingroup Innovation Foundation (VINIF) under project code VINIF.2023.DA.057. A part of this work was carried out while we were associate members of the Simons Astrophysics Group at ICISE (Quy Nhon, Viet Nam). We acknowledge the financial support from the Simons Foundation (No. 916424, N.H.) 
and the warm hospitality of the ICISE staff. Additionally, N.T.P. acknowledges financial support from the World Laboratory, N.T.T. acknowledges support through a Neighbourhood, Development and International Cooperation Instrument (NDICI) scholarship funded by the European Union in the framework of the Erasmus+, Erasmus Mundus Joint Master in Astrophysics and Space Science – MASS. Views and opinions expressed are, however, those of the author(s) only and do not necessarily reflect those of the European Union or the granting authority, European Education and Culture Executive Agency (EACEA). Neither the European Union nor the granting authority can be held responsible for them. This paper makes use of the following ALMA data: ADS/JAO.ALMA\#2015.1.00168.S, 2016.1.00484.L, and 2018.1.01198.S. ALMA is a partnership of ESO (representing its member states), NSF (USA), and NINS (Japan), together with NRC (Canada), NSTC and ASIAA (Taiwan), and KASI (Republic of Korea), in cooperation with the Republic of Chile. The Joint ALMA Observatory is operated by ESO, AUI/NRAO, and NAOJ. 
\clearpage
\appendix 
\section{Other molecular lines}
\label{sec:other_lines}
\counterwithin{figure}{section}
\setcounter{figure}{0}
Band 7 observations also include the molecular line emissions of HCO$^+$(4--3) and HCN(4--3), which trace features similar to those of CO and its isotopologues. The integrated intensity maps, brightness temperature maps, and velocity maps of these lines are shown in Figure \ref{fig:hcop+hcn}, while the PV diagrams cut along the major axis of the dust disk within the \asn~and \ass~components are shown in Figure \ref{fig:pv_hcop+hcn}.

\begin{figure*}[ht!]
    \centering
    \includegraphics[width=0.85\linewidth]{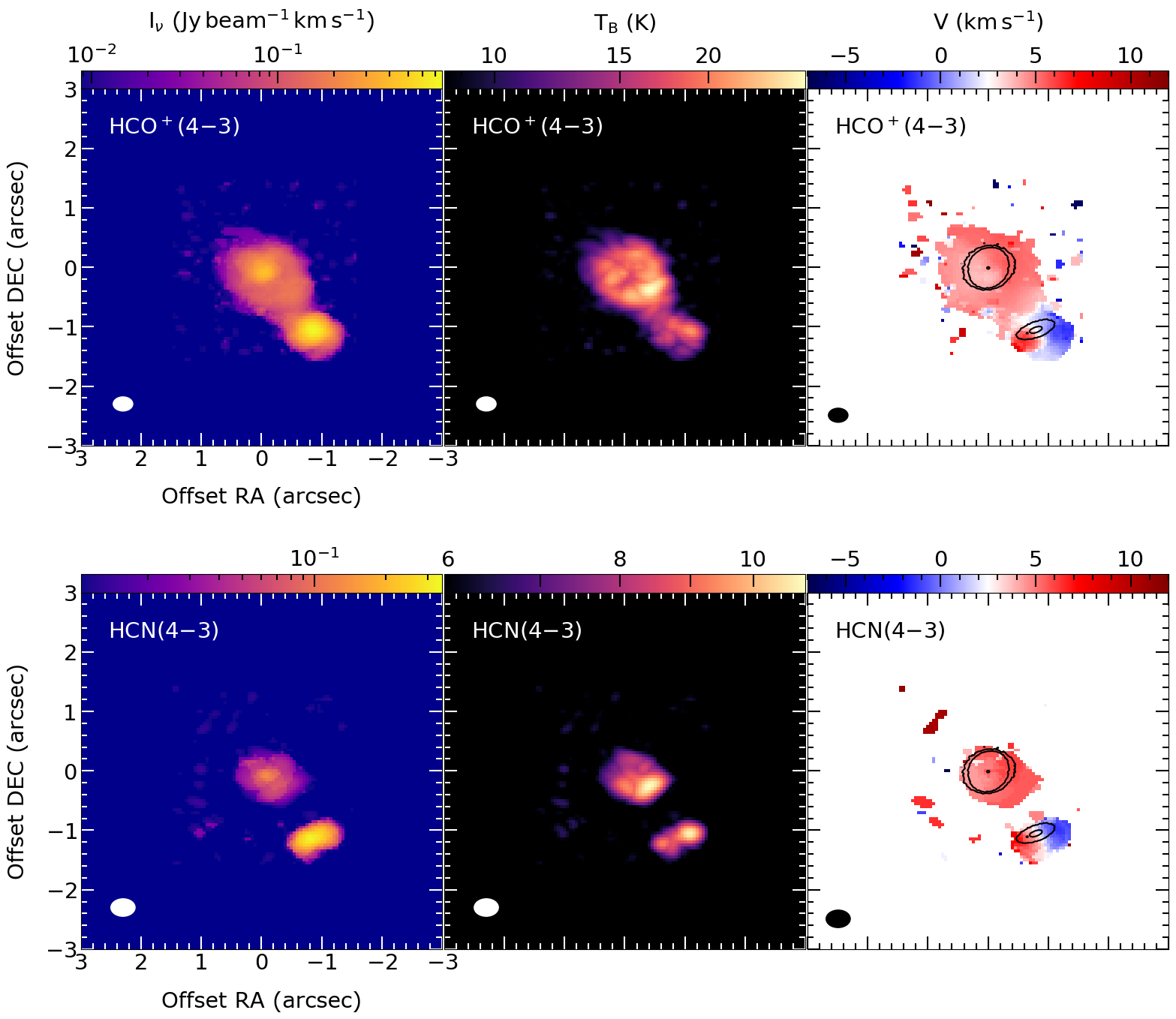}
    \caption{\textit{From left to right:} Integrated intensity, peak brightness and velocity maps of HCO$^+$(4--3) (\textit{top}) and HCN(4--3) (\textit{bottom}). The velocity maps of each molecular line are overlaid with the contours of dust emission at 1.3 mm, with the levels of [5, 25, 300]$\sigma$. The beams are shown in the lower left corner of each panel.} \label{fig:hcop+hcn}
\end{figure*}

\begin{figure*}[ht!]
    \centering
    \includegraphics[width=0.85\linewidth]{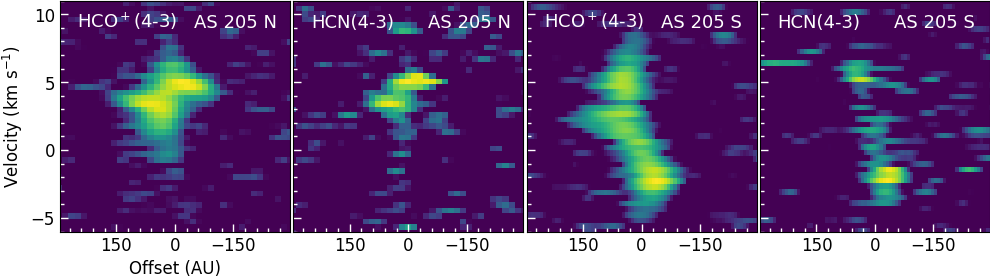}
    \caption{Position-Velocity diagrams cut along major axis dust disk toward AS 205 N (\textit{left panels})  and AS 205 S (\textit{right panels}) of the lines HCO$^+$(4--3) and HCN(4--3), respectively.} \label{fig:pv_hcop+hcn}
\end{figure*}

\section{Supplemental Figures}
\label{sec:optical_depth}
\begin{figure}[!h]
    \centering
    \includegraphics[width=0.75\linewidth]{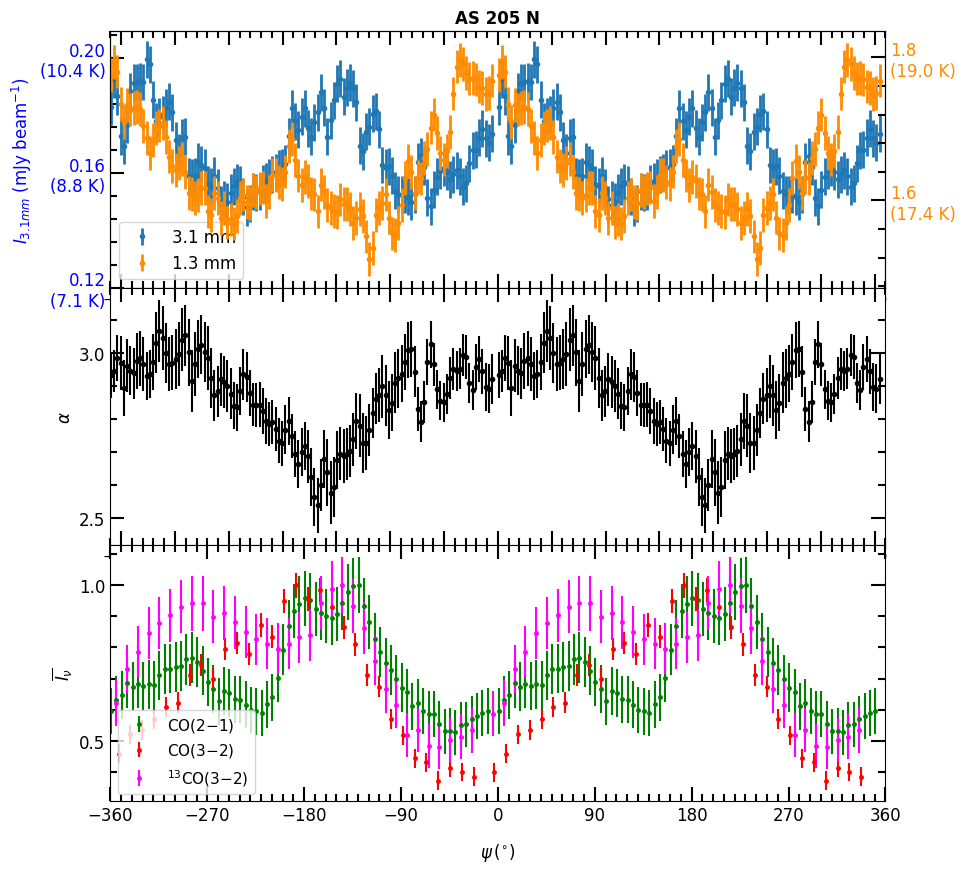}
    \caption{Same as left panel of Figure \ref{fig:alpha_AS205}, but with the azimuthal interval [$-360^{\circ},\,360^{\circ}$] in the abscissa.}    \label{fig:alpha_2}
\end{figure}

The azimuthal distribution of dust continuum intensity at 3.1 and 1.3 mm, their corresponding spectral index, and normalized integrated intensity for CO($2-1$), CO($3-2$), and $^{13}$CO($3-2$) in AS 205 N in the azimuthal interval [$-360^{\circ},\,360^{\circ}$] is shown in Figure \ref{fig:alpha_2}.

\clearpage
\bibliography{references}{}
\bibliographystyle{aasjournalv7}

\end{document}